# An Efficient Elliptic Curve Cryptography Arithmetic Using Nikhilam Multiplication

[1,] Prokash Barman, [2,] Banani Saha

[1, 2,] *Department of Computer Science & Engineering, University of Calcutta*
*Kolkata, India*

-----------------------------------------------------------**ABSTRACT**-----------------------------------------------------------
*Multiplication is one of the most important operation in Elliptic Curve Cryptography (ECC) arithmetic. For point addition and point doubling in ECC scalar (integer) multiplication is required. In higher order classical (standard) multiplication many intermediate operations are required. Reduced operation in multiplication will increase the functional speed of ECC arithmetic. These goals can be achieved using ancient multiplication algorithm namely Nikhilam Sutra. Nikhilam Sutra is one of the Sutra (algorithm) within 16 Vedic mathematics Sutras (algorithms). Nikhilam Sutra is efficient for multiplying two large decimal numbers. The Sutra reduces multiplication of two large numbers into two smaller numbers multiplication. The functional speed of Elliptic Curve Cryptography can be increased using Nikhilam method for scalar multiplication.*

## I. INTRODUCTION

Cryptography is a technique of making message secure or secret. Sensitive information can stored or transmitted across secure or insecure network with secure or secret message transmission method. So that an unauthorised person can't access the secret message. One of the best public key cryptographic methods to secure message is Elliptic Curve Cryptography. In this cryptographic method, major time consuming operations are **(1) Point Addition and (2) Point Doubling** which needs scalar (integer) multiplication.

Generally, multiplication uses standard or classical method. But classical method to multiply two n-bit integer requires **n²** operations [5]. Karatsuba method of multiplication uses divide-and-conquer technique to multiply two n-bit integers in **n^log₃** operations [5]. But for small inputs Karatsuba algorithm works slower than the classical multiplication algorithm because of recursive operational overhead. To overcome this problem we use Nikhilam Sutra of Vedic Mathematics. This Nikhilam Sutra (algorithm) performs large integer multiplication by converting it into small integer multiplication along with some addition and shift operation.

This paper is organised in the following sections. Section II describes Elliptic Curve Cryptography Arithmetic along with some conventional scalar multiplication methods. Nikhilam Navatascharamam Deshatah (Nikhilam Sutra) have been discussed in section III. The use of Nikhilam Sutra in ECC is depicted in section IV. Section V conclusion is described.

## II. ELLIPTIC CURVE CRYPTOGRAPHY ARITHMETIC

An Elliptic Curve is defined as an equation having set of solution with the point at infinity. The elliptic **y²+xy=x³+ax+b** in GF(2m) **and y²=x³+ax+b** in GM(P) are called weierstrass equations. Variables and coefficients are chosen from a large finite field. These points form a group. The group operation for elliptic curve cryptography are point multiplication, point addition and point doubling. Elliptic Curve Point Multiplication is the operation of successively adding a point along an elliptic curve to itself repeatedly- which is called as "**Elliptic Curve Scalar Multiplication**" [6]. The operation denoted as

**nP = P + P + P + P + . . . . . . . . . . . + P** for some scalar (integer) n and a point P(x,y) that lies on the curve **E: $y^2 = x^3 + ax + b$** (Weirerstrass Curve). The security of Elliptic Curve Cryptography depends on the intractability of determining n from **Q=nP**, given known value of **Q** and **P**. It is known as Elliptic Curve discrete logarithm problem.

**Point addition**

Point addition [7] is defined as taking two points along a curve E and computing where a line through them intersects the curve. The negative of the intersection point is used as the result of the addition.

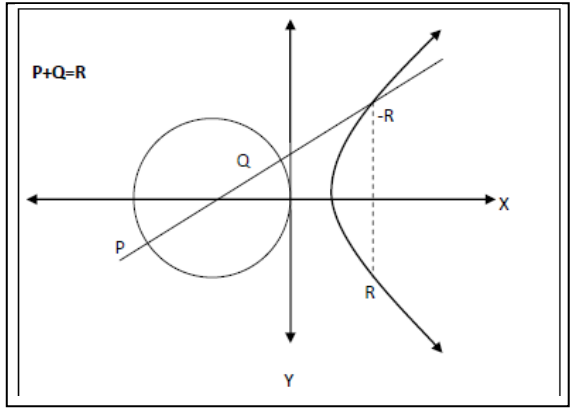

**ECC Point addition**

The operation is denoted as P + Q = R or $(X_p, Y_p) + (X_q, Y_q) = (X_r, Y_r)$. This can be algebraically calculated by :

$$\lambda = \frac{Y_q - Y_p}{X_q - X_p}$$
$$X_r = \lambda^2 - X_p - X_q$$
$$Y_r = \lambda(X_p - X_r) - Y_q$$

Note that we assume the elliptic field is given by $X^3 + aX + b$

**Point Doubling**

Point doubling [7] is similar to point addition, except one takes the tangent of a single point and finds the intersection with the tangent line.

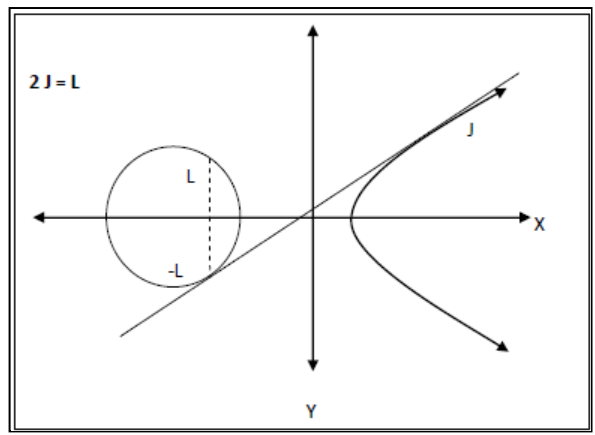

**ECC Point Doubling**

The operation is denoted as 2J=L or $(X_p, Y_p) + (X_p, Y_p) = (X_r, Y_r)$. This can be algebraically calculated by :

$$\lambda = \frac{3X_p^2 + a}{2Y_p}$$
$$X_r = \lambda^2 - 2X_p$$
$$Y_r = \lambda(X_p - X_r) - Y_q$$

Where a is the multiplication factor of X in the elliptic field given by $X^3 + aX + b$.

**Point multiplication [7] :**
The straight forward way of computing a point multiplication is through repeated addition. However, this is a fully exponential way to computing the multiplication. The multiplication method is as follows

**Double and add:**
To compute the double dP of a point P, start with the binary representation of d

$d = d_0 + 2d_1 + 2^2 d_2 + 2^3 d_3 + - - - - - - - - - 2^m d_m$

where $[d_0 - - - - - d_m] \Sigma \{1, 0\}$

**Algorithm:**
Input : P, m
Output: Q

Q=0
For i from m to 0 do
    Q=2Q (using point doubling)
    If $d_i = 1$ then Q = Q + P (using point addition)
Return Q

An alternative way of the above may also be derived using recursive function
**Algorithm:**
Input: $\int, P, n$
Output: $P + \int(P, n-1), \int(2P, n/2)$
$\int(P, n)$ is
   if n = 0 then return 0
   else if n mod 2=1 then
       $P + \int(P, n-1)$   [addition if n is odd]
   else $\int(2P, n/2)$ [doubling if n is even]

Here $\int is$ the function for doubling, P is the coordinate to double, n is the number of times to double the coordinate.

**Example:**
100P = $\int(P, 100)$ = 2 ( 2 ( P + 2 ( 2 ( P + 2P) ) ) )
This algorithm requires $Log_2(n)$ iteration of point doubling and addition to compute the full point multiplication. There are many variations of this algorithm such as window, sliding window, NAF, NAF-W, vector chains and Montogomery ladder which are beyond our current discussion.

### III. NIKHILAM NAVATASCHARAMAM DESHATAH (NIKHILAM SUTRA) MULTIPLICATION
One of the 16 Sutras of Vedic Mathematics is Nikhilam Sutra. It can be used to convert large digits multiplication to small digits multiplication with the help of few extra add, subtract and shift operations [5]. In some cases two-digit multiplication can be performed using only one digit multiplication instead of 3 one digit multiplication required with Karatsuba algorithm. The Vedic formula under consideration is Nikhilam Navatascharamam Deshatah [1], which means all from 9 and last from 10. The mathematical derivation of the algorithm is given as follows.

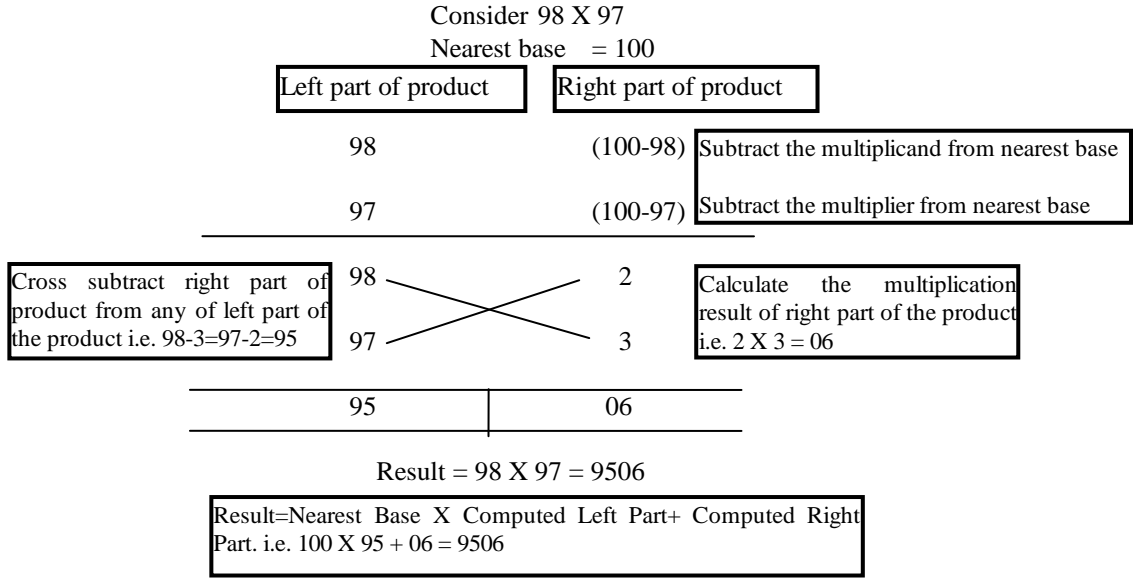

Figure 1: Multiplication using Nikhilam Sutra

With the help of Nikhilam Sutra two three digit multiplication can be performed with only one multiplication along with some addition and shift operations as shown in the table below

Table-I
Multiplication of 107 * 109

|  | Integer | Base Difference |
|---|---|---|
| Multiplicand | 107 | (100-107)= -7 |
| Multiplier | 109 | (100-109)= - 9 |
| Computation | (107-(-9))=(109-(-7))=116 | (-7)*(-9)=63 |
|  | 116 | 63 |
| Result | 11663 | |

For the above multiplication in standard method 9 multiplication is required and in Karatsuba algorithm 4 multiplication is required. The principle behind the Nikhilam Sutra is as follows:

Let the Multiplicand, m=x+a
    Multiplier, n=x+b

where, x is the nearest base, then  m*n= (x+a)*(x+b)=x(x+a+b) + ab

## IV. USE OF NIKHILAM MULTIPLICATION IN ECC SCALAR MULTIPLICATION

The point multiplication in ECC is basically includes Point doubling and Point addition operations as described in section II. These operations need scalar (integer) multiplication of large magnitude. In standard multiplication and Karatsuba multiplication method $n^2$ and $n^{\log_3}$ operations required to multiply two n digits numbers. Whereas the Nikhilam method require less multiplication operational steps. The use of Nikhlam method of multiplication in ECC scalar multiplication will increase the overall speed of Elliptic Curve Cryptography operation.

The use of scalar multiplication in ECC arithmetic is done in (1) Point Doubling and (2) Point Addition in the following cases

**Point Doubling:**

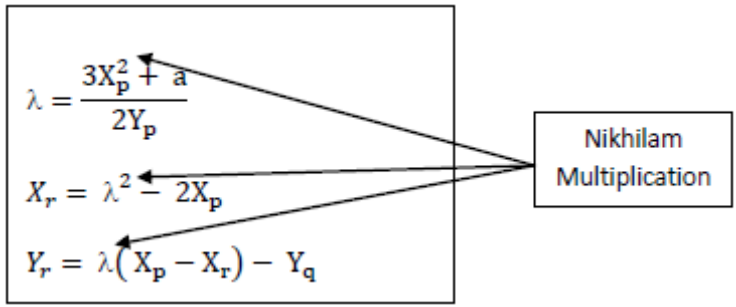

**Point addition**

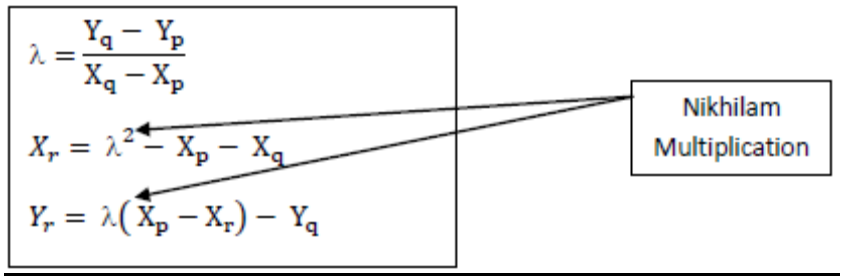

The actual scalar (integer) multiplication in ECC [7] is shown in the figures. In some cases squaring is also done with the nikhilam method.

## V. PROPOSED WORK

As Nikhilam multiplication needs less operational steps compared to other conventional multiplication methods, it is assumed that our proposed method of using nikhilam method in Point addition and Point doubling of ECC arithmetic will increase the operational speed also. To use the nikhilam multiplier in ECC we propose **Binary Multiplication** using Nikhilam method.

We can perform binary digit multiplication using Nikhilam sutra by converting n-bit multiplication to (n - 1)-bit multiplication and some additional add/subtract and shift operation. We can apply this conversion repeatedly until we get trivial multiplicand/multiplier or 1-bit multiplication. We can also put some threshold limit m where $1 < m < n$ up to which we would like to do this conversion.

2-bit multiplication can be performed using single 1-bit multiplication. For example if we have to multiply 11X11. Here multiplicand M = 11, and multiplier N = 11. We can proceed as follows:
1. Compute A = 11 - 10; Subtract the multiplicand from nearest base

2. Compute B = 11-10; Subtract the multiplier from the same base

3. Compute C = B * A = 1 * 1 = 1

4. Compute D = M + B = N + A = 11 + 1 = 100

5. Result 10 * D + C = 1001

The only multiplication required in this computation (Table II) is for C in step 3.

**TABLE II**
BINARY MULTIPLICATION OF 11*11

|             | Bits         | Base Difference |
|-------------|--------------|-----------------|
| Multiplicand | 11           | (11-10)=1       |
| Multiplier  | 11           | (11-10)=1       |
|             | (11+1)=100   | (1*1)=1         |
| Result      | 1001         |                 |

For 3-bit multiplication consider the example of 101 * 110

1. Compute A = 101*100; Subtract the multiplicand from nearest base
2. Compute B = 110 * 100; Subtract the multiplier from the same base
3. Compute C = 10 *1 = 10
4. Compute D = M + B = N + A = 101 + 10 = 111
5. Result 100 * D + C = 11110

**TABLE III**
BINARY MULTIPLICATION OF 101*110

|             | Bits           | Base Difference |
|-------------|----------------|-----------------|
| Multiplicand | 101            | (101-100)=1     |
| Multiplier  | 110            | (110-100)=10    |
|             | (101+10)=111   | (10*1)=10       |
| Result      | 11110          |                 |

In this computation also two 1-bit multiplication is performed. While in case of standard multiplication 9 multiplication is required, and Karatsuba algorithm use 4 multiplication.

## I. CONCLUSION

From the proposed binary multiplication method we can see that the Nikhilam multiplier uses less operational steps than the standard multiplier and other multiplier like Karatsuba algorithm. So it is found to be efficient in view of less operational steps and smaller multiplications. Hence the proposed binary multiplication method may be used in ECC efficiently. We are planning the hardware implementation of Elliptic Curve Cryptography system using binary Nikhilam multiplier embedded in FPGA.

## REFERENCES

[1] Jagadguru Swami Sri Bharath Krishna Tirthaji, "Vedic Mathematics or Sixteen Simple Sutras from Vedas", Motilal Bhandaridas, Varanasi (India), 1986.
[2] Bruce Schneier, "Applied Cryptography: Protocols, Algorithms and Source Code in C" – Second Edition.
[3] Mr. Dharmendra Madke, Prof. Sameena Zafar, "Polynomial Multiplication Using Karatsuba and Nikhilam Sutra"- International Journal of Advanced Research in Computer Science and Software Engineering (IJARCSSE), Vol-4, Issue-6, June-2014, pp. 1423-1428.
[4] Prokash Barman, Dr. Banani Saha, "E-Governance Security using Public Key Cryptography with special focus on ECC" – International Journal of Engineering Science Invention (IJESI), Vol-2, Issue-8, August 2013, pp. 10-16.
[5] Shri Prakash Dwivedi, "An Efficient Multiplication Algorithm Using Nikhilam Method"- arXiv:1307.2735v1 [cs.DS] 10 Jul 2013
[6] Krzysztof Jankowski, Pierre Laurent, Aidan O'Mahony, "Intel Polynomial Multiplication Instruction and its Usage for Elliptic Curve Cryptography" –White paper, April 2012
[7] Wikipedia, http://en.wikipedia.org/wiki/Elliptic_curve_point_multiplication - accessed on 17-09-2014